\documentclass[traditabstract]{aa} 
\usepackage{graphicx}
\usepackage{txfonts}
\usepackage{natbib,twoopt}
\usepackage[breaklinks=true]{hyperref} 
\bibpunct{(}{)}{;}{a}{}{,} 
\makeatletter
\newcommandtwoopt{\citeads}[3][][]{\href{http://adsabs.harvard.edu/abs/#3}%
{\def\hyper@linkstart##1##2{}%
\let\hyper@linkend\@empty\citealp[#1][#2]{#3}}}
\newcommandtwoopt{\citepads}[3][][]{\href{http://adsabs.harvard.edu/abs/#3}%
{\def\hyper@linkstart##1##2{}%
\let\hyper@linkend\@empty\citep[#1][#2]{#3}}}
\newcommandtwoopt{\citetads}[3][][]{\href{http://adsabs.harvard.edu/abs/#3}%
{\def\hyper@linkstart##1##2{}%
\let\hyper@linkend\@empty\citet[#1][#2]{#3}}}
\newcommandtwoopt{\citeyearads}[3][][]%
{\href{http://adsabs.harvard.edu/abs/#3}
{\def\hyper@linkstart##1##2{}%
\let\hyper@linkend\@empty\citeyear[#1][#2]{#3}}}
\makeatother
\def \lya\ {Lyman-$\alpha\,$}
\begin{document}
   \title{VLBI observations of the radio quasar J2228+0110 at $z$=5.95 and other field sources in multiple-phase-centre mode}

   \titlerunning{VLBI observations of J2228+0110 in the multiple-phase-centre mode}

   \authorrunning{H.-M. Cao et al.}

   \author{H.-M. Cao \inst{1, 2, 6}, S. Frey \inst{3}, L.I. Gurvits \inst{4, 5}, J. Yang \inst{4}, X.-Y. Hong \inst{1, 6}, Z. Paragi \inst{4}, A.T. Deller \inst{7}, \and \v{Z}. Ivezi\'{c} \inst{8}}

   \institute{Shanghai Astronomical Observatory, Chinese Academy of Sciences, Shanghai 200030,
              China. \email{hmcao@shao.ac.cn}
         \and
              University of Chinese Academy of Sciences, Beijing 100049, China
         \and
              F\"{O}MI Satellite Geodetic Observatory, PO Box 585, 1592 Budapest, Hungary
         \and
              Joint Institute for VLBI in Europe, Postbus 2, 7990 AA Dwingeloo, The Netherlands
         \and
              Department of Astrodynamics and Space Missions, Delft University of Technology,
              Kluyverweg 1, 2629 HS Delft, The Netherlands
         \and
              Key Laboratory of Radio Astronomy, Chinese Academy of Sciences, 210008 Nanjing, China
         \and
              ASTRON, The Netherlands Institute for Radio Astronomy, Postbus 2, 7990 AA Dwingeloo,
              The Netherlands
         \and
              Department of Astronomy, University of Washington, Seattle, Washington 98195, USA}

   \date{Received Dec 23, 2013; accepted Jan 24, 2014}

\abstract{A patch of sky in the SDSS Stripe 82 was observed at 1.6~GHz with Very Long Baseline Interferometry (VLBI) using the European VLBI Network (EVN). The data were correlated at the EVN software correlator at JIVE (SFXC). There are fifteen known mJy/sub-mJy radio sources in the target field defined by the primary beam size of a typical 30-m class EVN radio telescope. The source of particular interest is a recently identified high-redshift radio quasar J222843.54+011032.2 (J2228+0110) at redshift $z$=5.95. Our aim was to investigate the milli-arcsecond (mas) scale properties of all the VLBI-detectable sources within this primary beam area with  a diameter of 20$\arcmin$. The source J2228+0110 was detected with VLBI with a brightness temperature $T_{\rm b}$$>$$10^8$~K, supporting the active galactic nucleus (AGN) origin of its radio emission, which is conclusive evidence that the source is a radio quasar. In addition, two other target sources were also detected, one of them with no redshift information. Their brightness temperature values ($T_{\rm b}$$>$$10^7$~K) measured with VLBI suggest a non-thermal synchrotron radiation origin for their radio emission. The detection rate of 20\% is broadly consistent with other wide-field VLBI experiments carried out recently. We also derived the accurate equatorial coordinates of the three detected sources using the phase-referencing technique. This experiment is an early attempt of a wide-field science project with SFXC, paving the way for the EVN to conduct a large-scale VLBI survey in the multiple-phase-centre mode.}

\maketitle
\section{Introduction}
%
Very Long Baseline Interferometry (VLBI) surveys of extragalactic radio sources supply a wealth of information on the nature of physical phenomena in the close vicinity of active galactic nuclei (AGNs), with both astrophysical and cosmological applications \citep[for reviews, see e.g.][]{Gur04,Fre06}. However, traditional VLBI observations have very small fields of view compared to the average angular distance between compact extragalactic radio sources at the achievable sensitivity. Therefore, until recently, VLBI surveys are usually restricted to targeted observations of bright sources, and thus the growth of VLBI-detected AGN samples has been relatively slow over the past decades.

The high data recording rate of modern VLBI and long coherent on-source integration times achieved by the phase-referencing technique allow us to detect mJy-level or even sub-mJy-level weak sources. In its traditional way, phase-referencing involves regularly interleaving observations between the target source and a nearby, bright, and compact phase-reference calibrator \citep{Bea95}. The participating telescopes in a VLBI network change their pointing direction, nodding between the target and the reference source, with a period shorter than the atmospheric coherence scale. While VLBI surveys exploiting this technique are technically feasible \citep[e.g.][]{Gar99,Mos06,Fre08a}, they require a prohibitively large amount of VLBI observing resources to obtain deep mas-resolution radio imaging information on thousands of faint AGNs.

With the technique of in-beam phase-referencing \citep[e.g.][]{Gar01,Gar05,Len08}, any additional source that is present in the primary beam of the individual radio telescopes of the interferometer network can be phase-calibrated with a bright and compact primary object, without repointing the antennas. However, to observe a field as extended as possible by minimizing bandwidth and time-average smearing, experiments with a single correlation pass would require extremely high output data rates and thus produce very large data files, which are computationally challenging to analyse \citep[e.g.][]{Chi13}. Another possibility is to use as many different correlator passes as the number of targets, but then the required correlator resources are very considerable if not prohibitive.

The advent of multiple-phase-centre processing \citep{Del11,Mor11} using software VLBI correlators \citep[e.g.][]{Del07,Del11,Pid09} offers a possibility to produce datasets with many different phase centres located within the primary beam in a single correlator pass. The individual output file for each target has the usual averaging in time and frequency, and is easily manageable in the process of the calibration and imaging. Thus the in-beam phase referencing combined with the multiple-phase-centre correlation dramatically decreases the demand on the observing resources, as well as on computational requirements, while it only moderately increases the required correlator time. This in fact multiplies the surveying power of the VLBI network by a substantial factor. It has become feasible to plan and conduct systematic surveys of thousands of sources around suitable in-beam calibrators \citep{Del14}. This breakthrough will allow us to reveal mas-scale radio structures in a large sample of extragalactic sources, while pushing down the flux density limit by at least two orders of magnitude. In turn, this will bring an order of magnitude more VLBI-imaged extragalactic sources than are accessible for studies at present.

It is hard to predict all the novel science applications that will become possible with significantly (orders of magnitude) increased size samples of VLBI-imaged extragalactic radio sources. Relevant research fields include studies of the cosmological evolution of active galaxy populations, the role of star formation and accretion onto the central supermassive black hole, and their interplay (feedback mechanisms). Extensive optical sky surveys, such as the Sloan Digital Sky Survey \citep[SDSS,][]{Yor00} and the prospective surveys by the Large Synoptic Survey Telescope \citep[LSST,][]{Ive08}, reduce the problem of missing optical identifications and spectroscopic redshifts. High-resolution radio surveys can be matched with large multi-wavelength databases, such as in the infrared and X-ray bands, to study dust obscuration in the AGN host galaxies. Morphological classification of low-luminosity sources could eventually be made using a large sample. The properties of these sources can then be compared with those of well-studied powerful objects. Thousands of new radio AGNs imaged with VLBI and having measured cosmological redshift could provide the necessary impetus for renewed attempts to estimate cosmological model parameters using ``standard'' objects. Finally, collecting more AGNs with radio and optical identification could lead to a more accurate and dense astrometric link of the radio and optical reference frames.

At the 1.4-GHz flux densities $S_{1.4}$$>$1~mJy, the radio source population is believed to be dominated by radio-loud AGNs, and the star-forming galaxies start to become a major component of the sub-mJy population \citep[e.g.][]{Nor11}. Nevertheless, the exact fraction of AGNs among the sub-mJy radio sources is unknown. Using the multiple-phase-centre technique, \citet*{Mid11} observed an area containing the Chandra Deep Field South (CDFS) with the Very Long Baseline Array (VLBA), obtained a detection rate of 21\%, and identified seven new AGNs. The work, together with other wide-field experiments carried out recently \citep[e.g.][]{Mid13,Mor13}, demonstrated the capability of VLBI as a powerful tool for unveiling the nature of mJy/sub-mJy radio sources in multiple-phase-centre mode. In particular, this method can help identify the heavily obscured AGNs via their the radio emission that is completely free of dust obscuration.

The highest redshift ($z$$>$5.7) radio quasars belong to the mJy/sub-mJy radio source population due to their large luminosity distance. So far, three mJy-level $z$$\sim$6 quasars have been observed with VLBI \citep{Fre03,Fre05,Mom08,Fre08b,Fre11}. Their steep radio continuum spectra (spectral index $\alpha$$<$$-0.5$; $S_{\nu} \varpropto \nu^{\alpha}$, where $S_{\nu}$ is the flux density and $\nu$ the observing frequency) and compact radio structures ($<$$100$~pc) show that they have characteristics similar to the gigahertz-peaked-spectrum (GPS) sources known at lower redshifts. However, a larger sample is needed to draw a firm conclusion about whether most (if not all) of the highest redshift radio quasars are truly young radio sources in the early Universe.

The GPS and CSS (compact-steep-spectrum) sources are dominated by somewhat extended emission, although they both are confined to compact regions. The GPS sources have a size smaller than that of the narrow-line region (i.e. $<$1 kpc), while the CSS sources are entirely contained in their host galaxies (i.e. $<$15 kpc). They have a convex radio spectrum and barely show any variability in luminosity. They may be newly born radio sources that will eventually evolve into large Fanaroff--Riley \citep{Fan74} Class I (FR~I) or Class II (FR~II) radio sources (i.e. the youth scenario) or that are frustrated by the ambient dense gas/dust clouds \citep[i.e. the frustration scenario; see][for a review]{Dea98}. For the highest redshift GPS/CSS sources, both scenarios could be present. The very first generation of supermassive black holes had just recently developed their jets and expanding double mini-lobes, which at the same time were confined by the extremely dense interstellar medium (ISM) of their forming host galaxies \citep{Fal04}.

Recently, \citet*{Zei11} have identified the second radio-selected $z$$\sim$6 quasar \object{J222843.54+011032.2} (J2228+0110, hereafter) by matching the optical detections of the deep SDSS Stripe 82 and their radio counterparts in the Stripe 82 Very Large Array (VLA) Survey{\footnote{\url{http://sundog.stsci.edu/cgi-bin/searchstripe82}} \citep{Hod11}. The integrated flux density of $S_{1.4}$=1.32~mJy means that the source is the fourth $z$$\sim$6 radio-luminous quasar known to date with $S_{1.4}$$>$ 1~mJy and therefore is a feasible target for VLBI to probe the physical properties of the rare radio sources at the highest redshifts ($z$$\sim$6) at milli-arcsecond (mas) scales using the phase-referencing technique. In the project described in this paper, we observed J2228+0110 with the European VLBI Network (EVN) in the multiple-phase-centre mode, to simultaneously explore the mas-scale features of the neighbouring mJy/sub-mJy radio sources falling in the field of view (FOV).

In this paper, we adopt a flat $\Lambda$CDM cosmological model, with $H_0$=71~km\,s$^{-1}$\,Mpc$^{-1}$, $\Omega_{\rm m}$=0.27, and $\Omega_\Lambda$=0.73, which yields a projected linear scale of 5.86~pc\,mas$^{-1}$ at $z$=5.95. Our selection method for the field and the targets is described in Section~\ref{field}. The observations and the data reduction procedures are presented in Section~\ref{observations}. The results are shown in Section~\ref{results} and discussed in Section~\ref{discussion}. Some conclusions are drawn in Section~\ref{conclusions}.

\begin{table}
\caption{Fifteen target sources in the EVN experiment described in this paper.}
\begin{center}
\scalebox{0.9}{
\begin{tabular}{ccccc}
\hline
\hline
Source name & RA (J2000) & DEC (J2000) &  Flux density \\
            & h m s      & $\degr$ $\arcmin$ $\arcsec$ &  mJy   \\
\hline
 FIRST-6     &  22 28 42.206 &  01 14 47.73 &  1.06   \\
 J2228+0110\,* &  22 28 43.540 &  01 10 32.22 &  1.32  \\
 VLA-5\,*    &  22 28 49.842 &  01 13 37.47 &  1.76  \\
 NVSS-1      &  22 28 51.380 &  01 07 09.40 &  2.5    \\
 VLA-4\,*    &  22 28 51.452 &  01 12 03.39 &  0.37  \\
 NVSS-2      &  22 29 00.420 &  01 17 46.90 &  2.7    \\
 VLA-3\,*    &  22 29 18.059 &  01 06 18.43 &  7.97  \\
 VLA-1\,*    &  22 29 18.689 &  01 09 03.09 &  1.44  \\
 VLA-2\,*    &  22 29 25.487 &  01 17 52.61 &  0.43  \\
 FIRST-1     &  22 29 31.387 &  01 14 43.99 &  0.90   \\
 FIRST-2     &  22 29 33.802 &  01 15 38.08 &  2.02   \\
 FIRST-4     &  22 29 38.594 &  01 17 34.96 &  1.02   \\
 NVSS-3      &  22 29 38.790 &  01 18 05.20 &  3.1    \\
 FIRST-3     &  22 29 41.761 &  01 14 28.46 &  2.07   \\
 FIRST-5     &  22 29 52.917 &  01 12 19.57 &  0.92   \\
\hline
\end{tabular}}
\end{center}
\tablefoot{Col.~2--3: a priori right ascension (RA) and declination (DEC) in the J2000 equatorial coordinate system, which were used for the correlation; the coordinates of the sources marked with an asterisk are from the Stripe 82 VLA Survey \citep{Hod11}; the rest are from the FIRST \citep{Whi97} and NVSS \citep{Con98} surveys, as indicated by the source names; Col.~4: the integrated flux density of each source at 1.4~GHz measured in the respective low-resolution VLA observations mentioned above. \label{tbl-1}}
\end{table}
%
\section{Field and target selection}
\label{field}
%
We selected a target field containing the high-redshift quasar J2228+0110 and the phase calibrator J2229+0114 by placing the centre of the field (also the primary phase centre) midway between the two sources at RA=$22^{\rm h} 29^{\rm m} 17\fs67$ and DEC=$+01\degr 12\arcmin 44\farcs5$, so that phase-referencing observations could be conducted without nodding most of the EVN antennas, except the three largest ones we used (see below).

The additional target sources were searched within a radius of 10$\arcmin$ around the primary phase centre (P-centre) in the following databases: the Stripe 82 VLA Survey \citep{Hod11}, the Faint Images of the Radio Sky at Twenty-cm (FIRST) Survey \citep{Whi97} and the NRAO VLA Sky Survey \citep[NVSS,][]{Con98}. The search in these databases resulted in fifteen mJy/sub-mJy level radio sources for VLBI correlation (Table \ref{tbl-1}). Their approximate positions with respect to the P-centre are displayed in Fig. \ref{fig-1}.

The VLBA observations at 2.3 and 8.6~GHz \citep{Pet05} indicate that our phase calibrator J2229+0114 (RA=$22^{\rm h} 29^{\rm m} 51\fs8019$, DEC=$+01\degr 14\arcmin 56\farcs724$; the position RMS uncertainty is 0.54~mas)\footnote{\url{http://astrogeo.org/calib/search.html}} is just somewhat resolved with a total flux density $S_{2.3}$=194~mJy and a spectral index $\alpha$=$-0.15$$\pm$0.15 (assuming a 10\% flux density error). Such a strong and compact source is helpful for simplifying the phase calibration for the wide-field data.

\begin{figure*}
\centering
\includegraphics[width=13.cm]{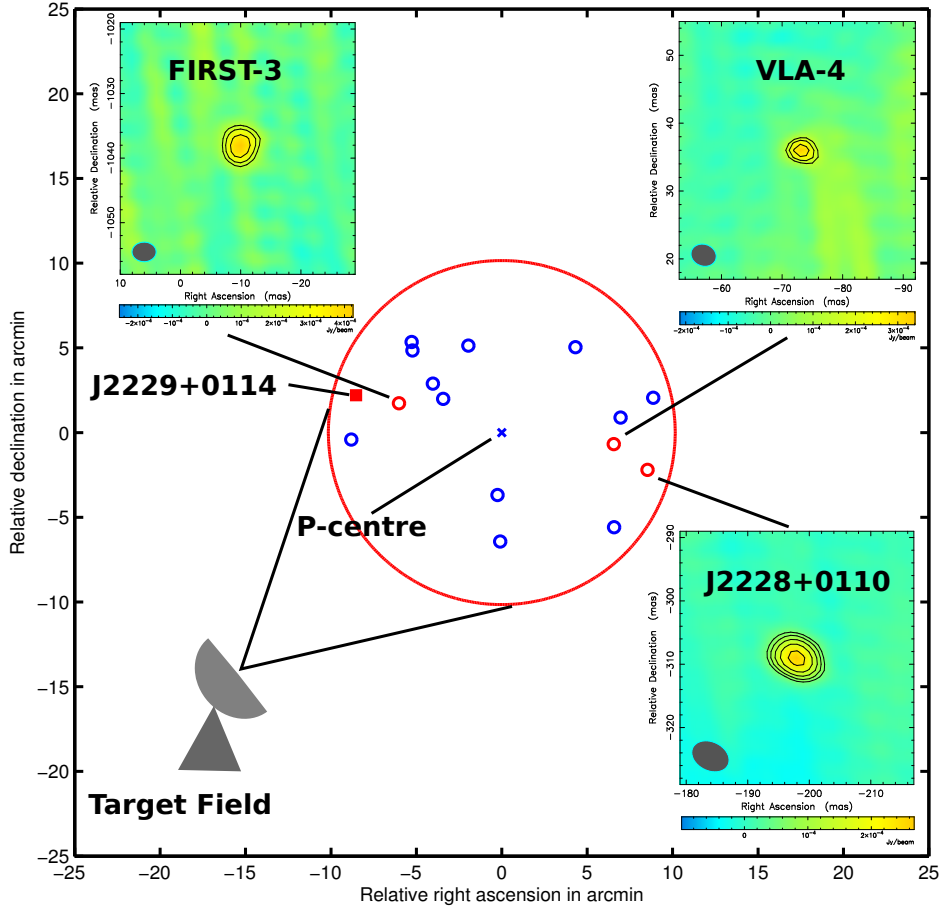}
\caption{EVN detection of the high-redshift quasar J2228+0110 and other two sources with the multi-phase-centre correlation technique. The label P-centre marks the pointing centre of the nine smaller telescopes working in the in-beam mode and the primary phase centre. The source J2229+0114 marked with a filled square served as the in-beam phase calibrator. The large circle is the primary beam size (FWHM) of a 32-m antenna at 1.6~GHz. Within the circle, there were 15 target sources observed. The naturally weighted images of the three sources detected in our EVN observations are shown in the insets. Here the coordinates are related to the a priori positions given in Table~\ref{tbl-1}. The first contour is drawn at 3$\sigma$ image noise, the positive contour levels increase by a factor of $\sqrt{2}$ in each image. The restoring beam size at FWHM is shown in the bottom-left corner of each inset. The image parameters are listed in Table~\ref{tbl-4}, where the difference in the synthesized beam sizes is also explained.
\label{fig-1}}
\end{figure*}
%
\section{EVN observations and data reduction}
\label{observations}
%
\subsection{Observations}
The EVN experiment took place at 1.6~GHz on 2011 Nov 1, in a combination mode of in-beam and nodding phase referencing \citep{Bea95, Gar05}. There were twelve antennas participating in the project, as listed in Table \ref{tbl-2}.
\begin{table}
\caption{The twelve antennas participating in the EVN project.}
\begin{center}
\scalebox{0.9}{
\begin{tabular}{ccc}
\hline
\hline
Telescope & Country & Diameter \\
          &         & m        \\
\hline
Jb        & United Kingdom  &  76      \\
Ef        & Germany &  100     \\
WSRT      & The Netherlands & 90\,* \\
On        & Sweden  & 25       \\
Mc        & Italy   & 32       \\
Tr        & Poland  & 32       \\
Bd, Sv, Zc & Russia & 32       \\
Sh, Ur    & China   & 25       \\
Hh        & South Africa & 26  \\
\hline
\end{tabular}}
\end{center}
\tablefoot{Telescope codes: Jb: Jodrell Bank Lovell; Ef: Effelsberg; WSRT: the Westerbork Synthesis Radio Telescope; On: Onsala85; Mc: Medicina; Tr: Toru\'{n}; Bd: Badary; Sv: Svetloe; Zc: Zelenchukskaya; Sh: Sheshan; Ur: Nanshan; Hh: Hartebeesthoek. $^*$ The WSRT was used in phased array mode with thirteen 25-m antennas; an equivalent diameter is given.
\label{tbl-2}}
\end{table}
The observations lasted for eight hours, recording at a data rate of 1024 Mbps with two circular polarizations, eight basebands per polarization and 16 MHz bandwidth per baseband. We employed two observing strategies. In the in-beam domain, the nine smaller antennas only pointed to the P-centre. Meanwhile, the three larger antennas (Ef, Jb, WSRT) observed J2228+0110 and J2229+0114 alternatively in the traditional nodding mode. This phase-referencing cycle was 5~min long, with 3~min spent on the high-redshift target J2228+0110. In the out-beam domain, all the telescopes acted synchronously, pointing to another phase calibrator at larger separation, J2226+0052, or to the fringe-finder source 3C454.3. With this configuration, we expected to achieve an on-axis theoretical thermal noise\footnote{EVN Calculator: \url{http://www.evlbi.org/cgi-bin/EVNcalc}} of $\sim$7~$\mu$Jy/beam for J2228+0110. For the other targets within the Effelsberg primary beam ($3\farcm25$ radius) around J2228+0110 and J2229+0114, the expected thermal noise was $\sim$10~$\mu$Jy/beam and $\sim$14~$\mu$Jy/beam, respectively, while for the rest it was $\sim$30~$\mu$Jy/beam.

The correlation of the recorded VLBI data was performed in the multiple-phase-centre mode at the EVN software correlator \citep[SFXC,][]{Pid09} of the Joint Institute for VLBI in Europe (JIVE), Dwingeloo, The Netherlands. The correlator output data were grouped into 16 multiple-source FITS-IDI files, each with a separate field source and all the out-beam calibrators, and averaged with an integration time of 2~s and 64 spectral channels per intermediate frequency channel (IF). The total size of the datasets for the post-processing was about 158 Gbytes.

\subsection{Data reduction}
We used a Gaussian model to approximate the antenna primary beam:
\begin{equation}
P(\Theta) = \exp \left( -\frac{\Theta^2}{2\sigma^2} \right),
\label{Eq-1}
\end{equation}
where $P(\Theta)$ is the relative power response, $\Theta$ is the angular distance of the source from the pointing direction of the antenna, and $\sigma$ is proportional to the full width at half maximum (FWHM) of the primary beam $\Theta_{1/2}$ as
\begin{equation}
\sigma^2 = \frac{\Theta_{1/2}^{2}}{8 \ln 2}.
\label{Eq-2}
\end{equation}
Here, $\Theta_{1/2}=K \times \lambda / D$, where $\lambda$ is the observing wavelength, and $D$ the aperture diameter. In the absence of detailed primary beam shape models for the individual EVN antennas, we adopted a small correction factor $K = 1.05$ reflecting that the apertures are not uniformly illuminated \citep[cf.][]{Wri11}. The antenna response of different apertures is shown in Fig.~\ref{fig-2}.
\begin{figure*}
\centering
\includegraphics[width=13.cm]{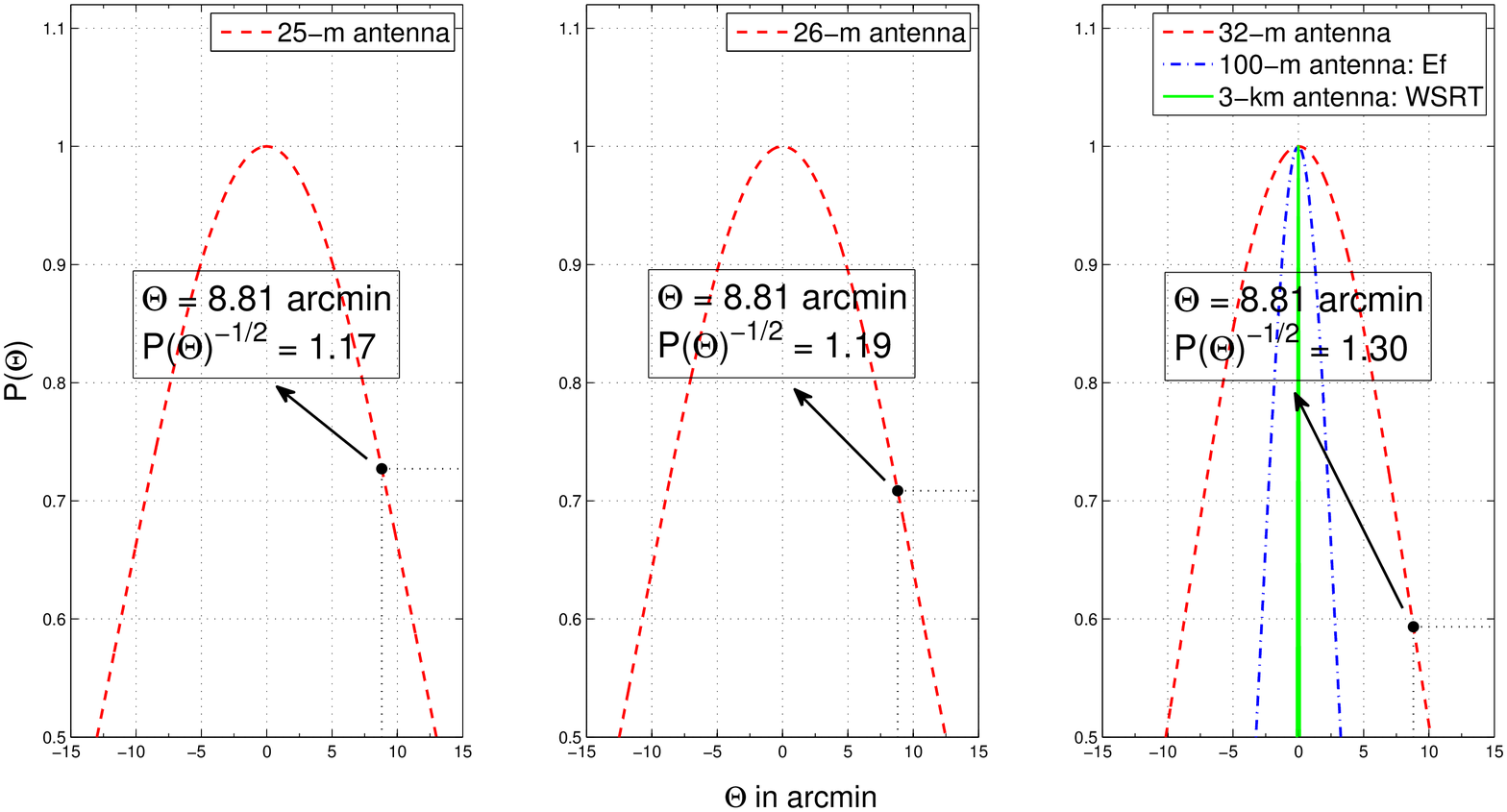}
\caption{The relative power response $P(\Theta)$ of radio telescopes with different aperture diameters. $\Theta$ is the angular distance between the source and the pointing (maximum response) direction of the antenna. The central wavelength ($\lambda$=18~cm) was used to estimate the FWHM of the primary beam. The primary beam correction values for J2229+0114 and J2228+0110 (at $\Theta=8\farcm81$) are also shown (see the text and Table \ref{tbl-3} for details). The WSRT in phased array mode has a collecting area equivalent to that of a single 90-m antenna, but its high angular resolution is determined by the maximum baseline length between the elements of the array (nearly 3~km).
\label{fig-2}}
\end{figure*}

The gain amplitude of each corresponding visibility was multiplied by the correction factors $P(\Theta)^{-1/2}$ of the two telescopes that form a given baseline. As an example, we list the primary beam correction values of different antenna apertures for J2229+0114 and J2228+0110 in Table~\ref{tbl-3}, and also plot them in Fig.~\ref{fig-2}. For the two datasets (J2229+0114 and J2228+0110), the three large telescopes (Ef, Jb, WSRT) do not need to be corrected because they operated in the nodding mode during the observations.
\begin{table}
\caption{Primary beam correction values for J2229+0114 and J2228+0110.}
\begin{center}
\scalebox{0.9}{
\begin{tabular}{cccc}
\hline
\hline
Antenna diameter & FWHM      &   Offset      &  Correction factor \\
$D$ (m)            & $\Theta_{1/2}$ ($\arcmin$) & $\Theta$ ($\arcmin$) & $P(\Theta)^{-1/2}$ \\
\hline
25               & 25.99     &   8.81           &   1.17     \\
26               & 24.99     &   8.81           &   1.19     \\
32               & 20.30     &   8.81           &   1.30     \\
\hline
\end{tabular}}
\end{center}
\tablefoot{The central wavelength ($\lambda$=18~cm) was used to estimate the FWHM of the primary beam. Offset is the angular distance between the source and the P-centre. \label{tbl-3}}
\end{table}

The in-beam reference source J2229+0114 was first calibrated and imaged. The primary beam correction was applied to the antenna gains by the task CLCOR of the NRAO Astronomical Image Processing System \citep[AIPS,][]{Gre03}\footnote{\url{http://www.aips.nrao.edu}}. A priori amplitude calibration was done in AIPS using the known antenna gain curves and the system temperatures measured regularly at the VLBI stations during the observations, or the nominal system equivalent flux density (SEFD) values for the three KVAZAR Network telescopes (Bd, Sv, Zc). Parallactic angle correction for the altitude-azimuth mounted antennas, and ionospheric correction were performed, too. The manual phase calibration and the global fringe-fitting have also been performed. The bandpass correction was carried out using the fringe-finder source 3C454.3. The visibility data of the calibrator source were  extracted and imaged in the Difmap program \citep{Shep97}. The antenna-based gain correction factors obtained after the first step of amplitude self-calibration were typically below 15\%.

A ParselTongue script \citep{Ket06} was then used to synchronously calibrate J2229+0114 and the 15 targets. The calibration procedure followed the steps described above. The antenna-based gain correction factors obtained in Difmap were fed back to AIPS and applied to the visibilities using the task SNCOR. The manual phase calibration and the global fringe-fitting were first performed for the calibrator dataset (i.e. the one containing J2229+0114). The component model of J2229+0114 obtained previously by the Difmap model-fitting routine was now taken into account, to compensate for small residual phases resulting from its non-point-like structure. The solution tables were then copied to each target dataset, where the phase calibration table was finally generated. The visibility data of each target source were extracted, and loaded into Difmap for imaging. The data from Jb had to be removed due to the improper phase calibration.

%
\section{Results}
\label{results}
%
A search for the VLBI detections was carried out in the naturally weighted dirty maps, each with 4$\arcsec$$\times$4$\arcsec$ in size. A peak brightness signal-to-noise ratio of 7 was used as a cutoff for a detection. The sub-field size is large enough to cover 3$\sigma$ a priori position errors, except for the NVSS sources (see discussion later in Sect.~\ref{discussion}). However, a larger area may also mean a higher probability for a ``phantom detection'' actually being a stochastic thermal noise peak. Three target sources (J2228+0110, VLA-4, and FIRST-3) were detected in the dirty image with a significance of 12.1$\sigma$, 7.7$\sigma$, and 9.3$\sigma$, respectively. The 7$\sigma$ upper limit for the peak brightness of the non-detections is $\sim$0.3--0.5~mJy/beam, depending on the actual sub-field and the number of contributing radio telescopes. For the detected sources, the naturally weighted images (Fig.~\ref{fig-1}) were made after model-fitting in Difmap. The image parameters are summarised in Table~\ref{tbl-4}, along with the detection limits for the other sources. The image noise achieved is higher than predicted because of the off-axis positions and the missing or flagged data from sensitive telescopes.
\begin{table}
\caption{The image parameters for the three detected sources in the top three rows (see Fig. \ref{fig-1}) and the brightness upper limits for the non-detections.}
\begin{center}
\scalebox{0.9}{
\begin{tabular}{ccccc}
\hline
\hline
Source name & Beam size         &   P.A.      & Peak br.        &  RMS br.      \\
            & mas $\times$ mas  &   $\degr$   & $\mu$Jy/beam    &  $\mu$Jy/beam \\
\hline
J2228+0110       & 6.34 $\times$ 4.56   &  65.5  &  267     &  19 \\
VLA-4            & 4.22 $\times$ 3.23   &  77.2  &  341     &  45 \\
FIRST-3          & 4.02 $\times$ 3.04   &  89.9  &  425     &  54 \\
\hline
FIRST-1          & 3.97 $\times$ 2.85   &  98.8  & $<$389   &  56 \\
FIRST-2          & 3.97 $\times$ 2.89   &  97.3  & $<$398   &  57 \\
FIRST-4          & 3.96 $\times$ 2.90   &  95.9  & $<$457   &  65 \\
FRIST-5          & 3.99 $\times$ 3.01   &  88.5  & $<$475   &  68 \\
FIRST-6          & 4.04 $\times$ 3.06   &  85.6  & $<$424   &  61 \\
NVSS-1           & 4.08 $\times$ 3.11   &  83.5  & $<$383   &  55 \\
NVSS-2           & 3.94 $\times$ 2.71   & 102.0  & $<$485   &  69 \\
NVSS-3           & 3.95 $\times$ 2.87   &  96.9  & $<$476   &  68 \\
VLA-1            & 3.97 $\times$ 2.75   & 101.8  & $<$430   &  61 \\
VLA-2            & 3.95 $\times$ 2.74   & 101.7  & $<$449   &  64 \\
VLA-3            & 3.94 $\times$ 2.70   & 102.5  & $<$491   &  70 \\
VLA-5            & 4.13 $\times$ 3.16   &  82.1  & $<$324   &  46 \\
\hline
\end{tabular}}
\end{center}
\tablefoot{Col.~1: the three detected sources (top) and the twelve undetected targets (bottom); Col.~2: restoring beam size at FWHM; Col.~3: position angle of the restoring beam major axis, measured from north through east; Col.~4: peak brightness for the detected sources (top) and 7$\sigma$ upper limits for the non-detections (bottom); Col.~5: off-source image noise (top) and the noise in the dirty image (bottom). The different beam size and image noise values are caused by missing or flagged antennas and/or different on-source times (see the text for details). \label{tbl-4}}
\end{table}
\subsection{J2228+0110}
We used one circular Gaussian brightness distribution model to fit the visibility data in Difmap. The VLBI-recovered flux density is 300$\pm$120~$\mu$Jy and the fitted source size is 1.71$\pm$0.47~mas (corresponding to a projected linear size of $\sim$10~pc).

The minimum resolvable size $\theta_{\rm min}$ of an interferometer for Gaussian brightness distribution in a naturally weighted image is given as
\begin{equation}
\theta_{\rm min}=b_{\psi} \sqrt{\frac{4\ln2}{\pi}\ln\left(\frac{\rm SNR}{\rm SNR-1}\right)},
\label{Eq-3}
\end{equation}
where $b_\psi$ is the restoring beam size at FWHM measured along an arbitrary position angle, and
${\rm SNR}$ the total flux density of the component divided by the image noise \citep[e.g.][]{Kov05}. The fitted source size is consistent with the $\theta_{\rm min}$ (1.30 mas) estimated with $b_\psi = \sqrt{b_{\rm max} \times b_{\rm min}}$, where $b_{\rm max}$ and $b_{\rm min}$ are the major and minor axis of the restoring beam, respectively, indicating that the source is practically unresolved in our EVN experiment.

The redshift-corrected brightness temperature for a circular Gaussian component \cite[e.g.][]{Con82} is expressed as
\begin{equation}
T_{\rm b}=1.22 \times 10^{12} (1+z)\frac{S_\nu}{\vartheta ^2\nu ^2},
\label{Eq-4}
\end{equation}
where $z$ is the redshift, $S_\nu$ the flux density of the component in Jy, $\vartheta$ the size of the circular Gaussian component at FWHM in mas, and $\nu$ the observing frequency in GHz. A VLBI detection with a brightness temperature exceeding $10^6$~K is interpreted as a signature of non-thermal synchrotron radiation, which can be used as an unambiguous AGN indicator, if the compact monochromatic radio luminosity\footnote{This quantity is in fact the power per unit frequency interval, and we call it power density, on the analogy of flux density, which is defined as the flux per unit frequency interval. Other terms widely used in the literature are (monochromatic) luminosity or power.} is $P_{\rm r} > 2 \times 10^{21}$~W\,Hz$^{-1}$ \citep[e.g.][]{Kew00,Mid11}. The inferred lower limit of the brightness temperature of $T_{\rm b}$$>$3.14$\times$$10^{8}$~K is fully consistent with the expectations for a
high-redshift radio quasar.

The phase-referenced coordinates (Table~\ref{tbl-5}) were derived using the AIPS verb TVMAXFIT. The position error is influenced by the thermal noise of the interferometer phases \citep{Fom99}, the position error of the in-beam phase calibrator J2229+0114, and the systematic error of phase-referencing observations. The last is the least important in our case because of the small angular distances involved.

\subsection{VLA-4 and FIRST-3}
The phased array of WSRT was flagged out from the other 14 target datasets owing to its small primary beam (Fig.~\ref{fig-2}). A circular Gaussian brightness distribution was used as a starting model to fit the measured visibilities of VLA-4 (J222851.44+011203.4) and FIRST-3 (J222941.76+011427.4). In the case of VLA-4, the radius of the Gaussian component converged toward zero, so we used a point source model instead. The corresponding model parameters are listed in Table~\ref{tbl-5}. The removal of data from two sensitive antennas (WSRT and Jb) in the central part of the network results in a smaller restoring beam size. The reason is the decrease in data weights at short spacings, hence the relative overweighting of the long baselines. The loss of data also results in an elevated image noise level compared to the case of J2228+0110. Because of the shorter on-source time, the image noise of FIRST-3 is higher than that of VLA-4 (Table~\ref{tbl-4}).

VLA-4 has an extended optical image, and the photometric redshift provided by the SDSS Data Release 9 (DR9)\footnote{\url{http://skyserver.sdss3.org/dr9}} database is 0.74$\pm$0.13, signifying the extragalactic nature of the object.
No obvious optical counterpart is found for FIRST-3 in the SDSS Stripe 82 database\footnote{\url{http://cas.sdss.org/stripe82}}. Since there is no redshift information available, we used $z$=0 to calculate the brightness temperature lower limit for FIRST-3 (Table \ref{tbl-5}).

Comparing the fitted size with the minimum resolvable angular size, FIRST-3 appears somewhat resolved. However, the more stretched structure of FIRST-3, which also accounts for the relatively larger position error, may be caused by the bandpass smearing effect of its large offset ($\sim$1$\arcsec$) with respect to the a priori phase centre (see Fig.~\ref{fig-1}, inset).

\begin{table*}[t]
\caption{Model parameters obtained for the three detected sources.}
\begin{center}
\scalebox{0.9}{
\begin{tabular}{ccccccccc}
\hline
\hline
Source name & Flux density & $\vartheta$ & $\theta_{\rm min}$ & $T_{\rm b}$ & RA (J2000) & DEC (J2000) & Pos. err.  \\
            & $\mu$Jy & mas & mas & $10^{8}$ K  & h m s      & $\degr$ $\arcmin$ $\arcsec$ &  mas \\
\hline
J2228+0110  & 300 $\pm$120 & 1.71$\pm$0.47 & 1.30 & $>$3.14 & 22 28 43.52679 & 01 10 31.9109 & 0.6 \\
VLA-4       & 340$\pm$190  & ...           & 1.30 & $>$1.56 & 22 28 51.44711 & 01 12 03.4259 & 0.6  \\
FIRST-3     & 990$\pm$420  & 3.92$\pm$1.48 & 0.78 & $>$0.28 & 22 29 41.76034 & 01 14 27.4219 & 1.0  \\
\hline
\end{tabular} }
\end{center}
\tablefoot{Col.~2: recovered by VLBI; the errors are estimated according to \citet*{Fom99}, assuming an additional 10\% error for flux density calibration and primary beam correction; Col.~3: size of the fitted circular Gaussian model component at FWHM; Col.~4: minimum resolvable angular size; Col.~5: estimated brightness temperature; Col.~6--7: astrometric position (right ascension and declination) measured by phase referencing; Col.~8: position error (1$\sigma$)
\label{tbl-5}}
\end{table*}
%
\section{Discussion}
\label{discussion}
%
The measured brightness temperature ($T_{\rm b}$$>$3.14$\times$$10^{8}$~K) of the high-redshift ($z$=5.95) source J2228+0110 and its compact radio power density ($P_{\rm 1.4}=8.3 \pm 3.3 \times10^{25}$~W\,Hz$^{-1}$ at 1.4~GHz in its rest frame) estimated from the VLBI flux density provide evidence for its identification as a radio quasar. Its radio-loudness parameter, the ratio of the total 5-GHz radio and 2500-$\AA$ optical flux density, is $R$$\approx$1100 \citep{Zei11}, indicating a highly radio-loud object\footnote{If the integrated VLA flux density of 1.32~mJy is used instead of the 0.31~mJy derived from the VLA peak brightness, $R$ becomes nearly 4700. While citing the results by \citet{Hod11} and \citet{Zei11}, we use the terminology that is different from theirs: we call ``peak brightness'' (measured in Jy/beam) what is apparently referred to in the cited papers as ``peak flux density'' and denominated in Jy. The description of this value in \citet{Zei11} indicates that this is indeed brightness.}. The rest-frame brightness temperature is expected to be $\sim$5$\times$$10^{10}$~K at the spectral turnover frequency of a self-absorbed system (e.g. the GPS/CSS sources), if we suppose that the particle and magnetic energy densities are in equipartition \citep[e.g.][]{Rea94}. Here the measured brightness temperature is significantly lower than the equipartition value, suggesting that the observing frequency (corresponding to 11.5~GHz in the source rest frame) may fall in the segment of the radio continuum with steep or inverted spectrum of a self-absorbed system.

The large discrepancy between our measured 1.6-GHz VLBI flux density of J2228+0110 ($S_{1.6}$=0.3~mJy) and its 1.4-GHz integrated VLA flux density \citep[$S_{1.4}$=1.32~mJy,][]{Hod11} implies that there may be significant extended structures resolved out in this source. In fact, some of the emission is already resolved out by the VLA ($1\farcs8$ scale) where the peak brightness is 0.31~mJy/beam \citep{Hod11}. While no discernible radio morphology in J2228+0110 is visible in the Stripe 82 VLA Survey image, as stated in the last sentence in Section 3.1 of \citet{Zei11}, the comparison of the integrated flux density (1.32~mJy) and the peak brightness (0.31 mJy/beam) reported in this survey indicate the existence of a radio structure on the arcsecond scale. The extended emission may originate in radio jets and lobes, and/or in the star-forming activity in the quasar host galaxy. However, a contribution of flux density variability cannot be ruled out either.

The 1.4-GHz radio continuum luminosity (i.e. power density) can be used as a star formation rate (SFR) indicator \citep[e.g.][]{Hop03}, which is justified by the tight radio--far infrared (FIR) correlation \citep[e.g.][]{Yun01}. Assuming that the difference between the integrated VLA flux density and the integrated VLBI flux density ($\sim$1~mJy) comes exclusively from star-forming activity, that is, from the radio supernovae/supernova remnants, we can estimate the SFR in the quasar host galaxy. Following the work of \citet*{Hop03}, we use
\begin{equation}
{\rm SFR}_{1.4} = \frac{L_{1.4}}{1.81 \times 10^{21} {\rm W\,Hz}^{-1}} \, {\rm M}_{\odot}\,{\rm yr}^{-1}, \label{Eq-5}
\end{equation}
where $L_{1.4}$ is the 1.4-GHz power density in W\,Hz$^{-1}$. For the K-correction, we assume a non-thermal spectral index of $\alpha$=$-0.8$ \citep[e.g.][]{Con92}, which is consistent with the typical spectral index values measured for the other known $z$$\sim$6 radio quasars \citep{Fre05,Fre08b,Fre11}. We obtain a SFR of $\sim$10$^5$~${\rm M}_{\odot}\,{\rm yr}^{-1}$. The implied extremely high SFR, exceeding the value of $\sim$1000~${\rm M}_{\odot}\,{\rm yr}^{-1}$ found in other $z$$\sim$6 quasars \citep{Ber03} by two orders of magnitude, indicates that difference between the VLA and VLBI flux densities cannot be explained by star formation alone, and significant AGN-related extended emission (jets/lobes) may exist in J2228+0110. These structures must remain partly unresolved with the VLA, but resolved out with the EVN, an they therefore extend to the angular scales of $\sim$100 mas to $\sim$1$\arcsec$, corresponding to $\sim$1--10~kpc in linear scale. Imaging observations using the intermediate baselines offered by the e-MERLIN array in the UK could possibly reveal these radio structures. The inferred linear size suggests that J2228+0110 is a CSS radio source. Once again, we note that we cannot rule out the possibility that the missing VLBI flux density is, at least partly, caused by the time-variable quasar emission or impact of scattering in the intervening propagation medium.

The VLBI flux density of the source VLA-4 appears to be consistent with what is measured by the VLA, which suggests that most of its radio emission comes from the AGN activity. Assuming the photometric redshift 0.74 and spectral index $-0.8$, its 1.4-GHz compact radio power density is in the order of $10^{23}$~W\,Hz$^{-1}$. However, in the absence of optical and any other data, the identification and the physical nature of the source FIRST-3 is not clear. It could be an obscured AGN at a relatively high redshift, possibly with intense star formation as the sub-millimetre galaxies \citep[e.g.][]{Bla02,Sch08} and, particularly so, if the star formation activity contributes an important fraction of the VLBI-resolved flux density. Multi-band observations are worth to be conducted to unveil its nature in the future.

With the EVN, we detected three of the targeted 15 radio sources known in the field (not counting here the bright in-beam calibrator, the 16th source observed in the field). While our detection rate of 20\% is generally consistent with the results of other recent multiple-phase-centre VLBI experiments, the exact comparison is complicated by the differences in the sensitivity levels and in the initial target selection methods and catalogues. For example, \citet*{Mid11} used the VLBA at 1.4~GHz to search for compact radio sources in the CDFS with a $\sim$0.5~mJy detection limit, based on a deep image made previously with the Australia Telescope Compact Array (ATCA). Their faintest target source had an ATCA flux density of 0.1~mJy, nearly four times lower than for the faintest one in our list (VLA-4; see Table~\ref{tbl-1}). Our inhomogeneous initial sample was drawn from three different catalogues, all made with the VLA at 1.4~GHz but at various epochs and, importantly, in different array configurations. This led to different angular resolutions, ranging from $1\farcs8$ for the Stripe 82 VLA Survey \citep{Hod11} to 45$\arcsec$ for the NVSS \citep{Con98}.
Moreover, with the typical 1$\sigma$ position uncertainty of about $4\farcs5$ for an unresolved NVSS source of the
flux density around 3~mJy \citep{Con98}, it was perhaps too optimistic to expect a VLBI detection within a map area of 4$\arcsec$$\times$4$\arcsec$. The three NVSS sources in our sample have no obvious FIRST counterparts, possibly indicating their resolved structures on angular scales of arcseconds. Indeed, no NVSS-only source
has been detected in our EVN experiment. We consider this as predictable and strong advice not to rely on NVSS-only targets in future VLBI experiments that are similar to the one described here. With the exclusion of the three NVSS-only targets in our sample, the overall detection rate in our experiment becomes 25\%.

The overall 21\% detection rate of \citet*{Mid11} is remarkably similar to ours. However, in the 1~mJy$\leq$$S_{1.4}$$\leq$10~mJy range, where the majority of our targets (11 of 15) fall, their detection rate is nearly 70\%. We detected only two sources (J2228+0110, FIRST-3) in this category. On the other hand, we could detect one of our four sub-mJy sources (VLA-4), while \citet*{Mid11} achieved 7\% in this flux density range. Since the number of sources involved in our experiment is very small, the detection percentages do not have statistical significance. With this caveat, we must note that our detection rate for the sources above 1~mJy seems to be lower than the 60--70\% expected from other recent studies \citep[e.g.][]{Nor11,Mid13}.

In their wide-field VLBA observations of M\,31 at 1.6~GHz, \citet*{Mor13} targeted 217 sources and report 16 firm detections, all background AGNs. These numbers are not easily compared to ours since the majority of their VLA-detected target sources actually belong to the M\,31 and are typically associated with compact H{\sc II} regions. In the absence of any nearby galaxy in the foreground, \citet*{Mid13} studied another 217 radio sources in the Lockman Hole/XMM field with mosaiced wide-field VLBA observations at 1.4~GHz and detected 65 of them (30\%). The typical detection threshold was 0.36~mJy/beam. They conclude that even in the flux density range 0.1~mJy$\leq$$S_{1.4}$$\leq$1~mJy, at least 15--25\% of the radio sources show AGN activity. Our result is consistent with this conclusion at the achieved (rather low) level of statistical confidence.

The most recent results of \citet*{Del14} are based on a sample of over 21\,000 FIRST target sources, much larger than studied in  previous surveys. The $\sim$4300 VLBA detections at 1.4~GHz, with a detection sensitivity of $\sim$1~mJy/beam, allowed the authors to investigate the detection fraction as a function of the arcsecond-scale FIRST flux density and the compactness ratio defined as the ratio of the VLBI peak brightness to the FIRST peak brightness. In general, \citet*{Del14} found that over half of the arcsecond-scale radio sources contain compact structure on mas scales, but this rate depends on the arcsecond-scale flux density. In the 1~mJy$\leq$$S_{1.4}$$\leq$2~mJy range where our FIRST target sources are typically found, \citet*{Del14} give nearly a 50\% detection rate. The majority of their detections are at the compactness ratios between 0.32 and 0.64, where our only detected FIRST target (FIRST-3, with a compactness ratio of about 0.5) is also located. Once again, we stress that any detection fraction derived from our small sample is statistically insignificant.
%
\section{Conclusions}
\label{conclusions}
%
We observed and detected the second radio-selected $z$$\sim$6 quasar J2228+0110 with the EVN in wide-field mode, using in-beam phase referencing. In addition, with the same observing data, we applied the novel multiple-phase-centre correlation technique and detected two other target radio sources within a 20$\arcmin$-diameter field. The measured brightness temperature of J2228+0110 is similar to that of the three other known $z$$>$5.7 radio quasars, and the inferred size suggests it might be a CSS source. The high brightness temperature of another detected object, VLA-4 (\object{J222851.44+011203.4}), together with the photometric redshift of 0.74$\pm$0.13, implies AGN activity in its host galaxy. The nature of FIRST-3 (\object{J222941.76+011427.4}) remains unclear in the absence of optical identification and redshift information. It is most likely an obscured AGN. The accurate astrometic coordinates of the three targets were also derived using the phase-referencing technique. The detection rate of 20\% is broadly consistent with the findings of larger surveys, but is lower than expected for the sub-sample with $S_{1.4}$$>$1~mJy. The result confirms that an important fraction of mJy/sub-mJy radio sources are (at least partially) powered by AGN.

With the project presented here, and while successfully imaging a rare high-redshift radio quasar, we demonstrated the ability of the EVN to obtain high-resolution structural information on unrelated radio sources that happen to lie within $\sim$10$\arcmin$ of the primary phase centre. Experiments like this could later be used for exploring the close vicinity of any strong calibrator source observed with the VLBI network for sufficiently long time, irrespective of the main science goal. This approach does not place any special requirement on VLBI array pointing and scheduling and, in particular, does not require any excess observing time from the VLBI network. Therefore the data can be obtained very efficiently, in piggyback mode. The only extra effort needed is the multiple-phase-centre correlation of the recorded data. In this paper, we demonstrated how to analyse the data from such an experiment.
\begin{acknowledgements}
We thank the anonymous referee for helping us improve the presentation of our results.
H.-M. C. thanks M. K. Argo for helpful advice on wide-field imaging.
We acknowledge support from the Royal Dutch Academy of Sciences (KNAW), the Chinese Academy of
Sciences (CAS; project no. 10CDP005), the Opening Project of the Shanghai Key Laboratory of Space Navigation and
Position Techniques (Grant No. 13DZ2273300), and the Hungarian Scientific Research Fund (OTKA K104539). The
EVN is a joint facility of European, Chinese, South African, and other radio astronomy institutes funded by
their research councils. Funding for the SDSS and SDSS-II has been provided by the Alfred P. Sloan Foundation,
the Participating Institutions, the National Science Foundation, the U.S. Department of Energy, the National
Aeronautics and Space Administration, the Japanese Monbukagakusho, the Max Planck Society, and the Higher
Education Funding Council for England. The SDSS Web Site is \url{http://www.sdss.org/}.

\end{acknowledgements}

\begin{thebibliography}{aa}
%
\bibitem[Beasley \& Conway (1995)]{Bea95} Beasley, A.J., \& Conway, J.E. 1995, in Very Long Baseline Interferometry and the VLBA, ed. J.A. Zensus, P.J. Diamond, \& P.J. Napier (San Francisco: ASP), ASP Conf. Ser., 82, 327

\bibitem[Bertoldi et al. (2003)]{Ber03} Bertoldi, F., Carilli, C.L., Cox, P., et al. 2003, \aap, 406, L55

\bibitem[Blain et al. (2002)]{Bla02} Blain, A.W., Smail, I., Ivison, R.J., Kneib, J.P, \& Frayer, D.T. 2002, Phys. Rep., 369, 111

\bibitem[Chi et al. (2013)]{Chi13} Chi, S., Barthel, P.D., \& Garrett, M.A. 2013, \aap, 550, A68

\bibitem[Condon et al. (1992)]{Con92} Condon, J.J. 1992, \araa, 30, 575

\bibitem[Condon et al. (1982)]{Con82} Condon, J.J., Condon, M.A., Gisler, G., \& Puschell, J.J. 1982, \apj, 252, 102

\bibitem[Condon et al. (1998)]{Con98} Condon, J.J., Cotton, W.D., Greisen, E.W., et al. 1998, \aj, 115, 1693

\bibitem[Deller et al. (2007)]{Del07} Deller, A.T., Tingay, S.J., Bailes, M., \& West, C. 2007, \pasp, 119, 853

\bibitem[Deller et al. (2011)]{Del11} Deller, A.T., Brisken, W.F., Phillips, C.J., et al. 2011, \pasp, 123, 275

\bibitem[Deller \& Middelberg (2014)]{Del14} Deller, A.T., \& Middelberg, E. 2014, \aj, 147, 14

\bibitem[Falcke et al. (2004)]{Fal04} Falcke, H., K\"{o}rding, E., \& Nagar, N.M. 2004, NewAR, 48, 1157	

\bibitem[Fanaroff \& Riley (1974)]{Fan74} Fanaroff, B.L., \& Riley, J.M. 1974, \mnras, 167, 31P	
	
\bibitem[Fomalont (1999)]{Fom99} Fomalont, E.B. 1999, in Synthesis Imaging in Radio Astronomy II, ed. G.B. Taylor, C.L. Carilli, \& R.A. Perley (San Francisco: ASP), ASP Conf. Ser., 180, 301

\bibitem[Frey (2006)]{Fre06} Frey, S. 2006, Proceedings of Science, PoS (8thEVN) 001

\bibitem[Frey et al. (2003)]{Fre03} Frey, S., Mosoni, L., Paragi, Z., \& Gurvits, L.I. 2003, MNRAS, 343, L20

\bibitem[Frey et al. (2005)]{Fre05} Frey, S., Paragi, Z., Mosoni, L., \& Gurvits, L.I. 2005, \aap, 436, L13

\bibitem[Frey et al. (2008a)]{Fre08a} Frey, S., Gurvits, L.I., Paragi, Z., et al., 2008a, \aap, 477, 781

\bibitem[Frey et al. (2008b)]{Fre08b} Frey, S., Gurvits, L.I., Paragi, Z., \& Gab\'anyi, K.\'E. 2008b, \aap, 484, L39

\bibitem[Frey et al. (2011)]{Fre11} Frey, S., Paragi, Z., Gurvits, L.I., Gab\'anyi, K.\'E., \& Cseh, D. 2011, \aap, 531, L5

\bibitem[Garrett et al. (2001)]{Gar01} Garrett, M.A., Muxlow, T.W.B., Garrington, S.T. et al. 2001, \aap, 366, L5

\bibitem[Garrett et al. (2005)]{Gar05} Garrett, M.A., Wrobel, J.M., \& Morganti, R. 2005, \apj, 619, 105

\bibitem[Garrington et al. (1999)]{Gar99} Garrington, S.T., Garrett, M.A., \& Polatidis, A. 1999, NewAR, 43, 629

\bibitem[Greisen (2003)]{Gre03} Greisen, E.W. 2003, in Information Handling in Astronomy -- Historical Vistas, ed. A. Heck, Astrophys. Space Sci. Lib., 285, 109

\bibitem[Gurvits (2004)]{Gur04} Gurvits, L.I. 2004, NewAR, 48, 1211

\bibitem[Hodge et al. (2011)]{Hod11} Hodge, J.A., Becker, R.H., White, R.L., Richards, G.T., \& Zeimann, G.R. 2011, \aj, 142, 3

\bibitem[Hopkins et al. (2003)]{Hop03} Hopkins, A.M., Miller, C.J., Nichol, R.C., et al. 2003, \apj, 599, 971

\bibitem[Ivezi\'{c} et al. (2008)]{Ive08} Ivezi\'{c}, \v{Z}., Tyson, J.A., Acosta, E., et al., 2008, arXiv:0805.2366

\bibitem[Kewley et al. (2000)]{Kew00} Kewley, L.J., Heisler, C.A., Dopita, M.A., et al. 2000, \apj, 530, 704

\bibitem[Kettenis et al. (2006)]{Ket06} Kettenis, M., van Langevelde, H.J., Reynolds, C., \& Cotton, B. 2006, in Astronomical Data Analysis Software and Systems XV, ed. C. Gabriel, C. Arviset, D. Ponz, \& S. Enrique (San Francisco: ASP), ASP Conf. Ser., 351, 497

\bibitem[Kovalev et al. (2005)]{Kov05} Kovalev, Y.Y., Kellermann, K.I., Lister, M.L., et al. 2005, \aj, 130, 2473

\bibitem[Lenc et al. (2008)]{Len08} Lenc, E., Garrett, M.A., Wucknitz, O., Anderson, J.M. \& Tingay, S.J. 2008, \apj, 673, 78

\bibitem[McGreer et al. (2006)]{McG06} McGreer, I.D., Becker, R.H., Helfand, D.J., \& White, R. L. 2006, \apj, 652, 157

\bibitem[Middelberg et al. (2011)]{Mid11} Middelberg, E., Deller, A., Morgan, J., et al. 2011, \aap, 526, 74

\bibitem[Middelberg et al. (2013)]{Mid13} Middelberg, E., Deller, A.T., Norris, R.P., et al. 2013, \aap, 551, 97	

\bibitem[Momjian et al. (2008)]{Mom08} Momjian, E., Carilli, C.L., \& McGreer, I.D. 2008, \aj, 136, 344

\bibitem[Morgan et al. (2011)]{Mor11} Morgan, J.S., Mantovani, F., Deller, A.T., et al. 2011, \aap, 526, 140

\bibitem[Morgan et al. (2013)]{Mor13} Morgan, J.S., Argo, M.K., Trott, C.M., et al. 2013, \apj, 768, 12

\bibitem[Mosoni et al. (2006)]{Mos06} Mosoni, L., Frey, S., Gurvits, L.I., et al. 2006, \aap, 445, 413

\bibitem[Norris et al. (2011)]{Nor11} Norris, R.P., Hopkins, A.M., Afonso, J., et al. 2011, \pasa, 28, 215

\bibitem[O'Dea (1998)]{Dea98} O'Dea, C.P. 1998, \pasp, 110, 493

\bibitem[Petrov et al. (2005)]{Pet05} Petrov, L., Kovelev, Y.Y., Fomalont, E., \& Gordon, D. 2005, \aj, 129, 1163

\bibitem[Pidopryhora et al. (2009)]{Pid09} Pidopryhora, Y., Keimpema, A., \& Kettenis, M. 2009, Proceedings of Science, PoS (EXPReS09) 046

\bibitem[Readhead (1994)]{Rea94} Readhead, A.C.S. 1994, \apj, 426, 51

\bibitem[Schinnerer et al. (2008)]{Sch08} Schinnerer, E., Carilli, C.L., Capak, P., et al., 2008, \apj, 689, L5

\bibitem[Shepherd (1997)]{Shep97} Shepherd, M.C. 1997, in Astronomical Data Analysis Software and Systems VI, ed. G. Hunt, \& H.E. Payne (San Francisco: ASP), ASP Conf. Ser., 125, 77

\bibitem[White et al. (1997)]{Whi97} White, R.L., Becker, R.H., Helfand, D.J., \& Gregg, M.D. 1997, \apj, 475, 479

\bibitem[Wrigley (2011)]{Wri11} Wrigley N. 2011, High Resolution Wide-Field Radio Imaging of the
GOODS North Field, MSc thesis (Univ. of Manchester)

\bibitem[York et al. (2000)]{Yor00} York, D.G., Adelman, J., Anderson, J.E., et al. 2000, \aj, 120, 1579

\bibitem[Yun et al. (2001)]{Yun01} Yun, M.S., Reddy, N.A., \& Condon J.J. 2001, \apj, 554, 803

\bibitem[Zeimann et al. (2011)]{Zei11} Zeimann, G.R., White, R.L., Becker, R.H., et al. 2011, \apj, 736, 57

%
\end{thebibliography}
\end{document}